\documentclass[12pt,a4paper]{amsart}
\usepackage[dvips,bookmarks,colorlinks=true,linkcolor=blue,urlcolor=red]{hyperref}
\usepackage{geometry}
\usepackage{latexsym,amsmath,amssymb,amsthm,color}
\newcommand{\Q}{{\mathbb Q}}
\newcommand{\C}{{\mathbb C}}

\newcommand{\R}{{\mathbb R}}
\newcommand{\Z}{{\mathbb Z}}
\newcommand{\N}{{\mathbb N}}

\newcommand{\dsize}{\textstyle}

\newcommand{\demo}{{\bf Proof.} \ }

\theoremstyle{plain}
\newtheorem{Th}{Theorem}
\newtheorem{Le}{Lemma}

\newtheorem{Cor}{Corollary}
\theoremstyle{definition}
\newtheorem{Rem}{Remark}

\title{Pointwise existence of the Lyapunov exponent for a
  quasi-periodic equation}
\author{Alexander Fedotov} \author{Fr{\'e}d{\'e}ric Klopp}
\address[Alexander Fedotov]{Departement of Mathematical Physics, St
  Petersburg State University, 1, Ulia\-novskaja, 198904 St
  Petersburg-Petrodvorets, Russia}
\email{\href{mailto:fedotov.s@mail.ru}{fedotov.s@mail.ru}}
\address[Fr{\'e}d{\'e}ric Klopp]{LAGA, Institut Galil{\'e}e, U.R.A 7539 C.N.R.S,
  Universit{\'e} de Paris-Nord, Avenue J.-B.  Cl{\'e}ment, F-93430
  Villetaneuse, France\\ and \\ Institut Universitaire de France}
\email{\href{mailto:klopp@math.univ-paris13.fr}{klopp@math.univ-paris13.fr}}
\thanks{Both authors acknowledge support of the CNRS and RFBR through
 the PICS grant n$^\circ $ 07-01-92169.}
\begin{document}
\maketitle
\section{Introduction}
\label{sec:introduction}
\subsection{Quasi-periodic finite difference equations}
\label{sec:quasi-peri-finite}
Consider the finite difference Schr{\"o}din\-ger equation
\begin{equation}
  \label{eq:1}
  (H_\theta\psi)(n)=\psi(n+1)+\psi(n-1)+v(n\omega+\theta)\psi(n)=E\psi(n)
\end{equation}
where $v:\R\to\R$ is continuous and periodic, $v(x+1)=v(x)$, \
$0<\omega<1$ and $0\le \theta<1$.
\\
When $\omega\not\in\Q$, the mapping $n\mapsto v(n\omega+\theta)$ is
quasi-periodic.
\\\\
The spectral theory of such quasi-periodic equations is very rich, and
the study has generated a vast literature; among the authors are
A.~Avila, Y.~Avron, J.~Bellissard, J.~Bourgain, V.~Buslaev,
V.~Chulaevsky, D.~Damanik, E.~Dinaburg, H.~Eliasson, A.~F.,
B.~Helffer, M.~Hermann, S.~Jitomirskaya, F.~K., R.~Krikorian, Y.~Last,
L.Pastur, J.~Puig, M.~Shubin, B.~Simon, Y.~Sina{\"\i}, J.~Sj{\"o}strand,
S.~Sorets, T.~Spencer, M.~Wilkinson and many others (see
e.g.~\cite{Pu:04} for a recent survey).
\\
Speaking about intriguing spectral phenomena, one can mention for
example that 
%
%%% Comme la relation entre la nature spectrale et les proprietes de
%%%  $\omega$ est connue pour une classe large de $v$, j'ai coupe
%%%   {\itemize} en deux : 

\begin{itemize}
\item for such equations, the spectral nature depends on 
  the ``number theoretical'' properties of the frequency $\omega$;
\end{itemize}
and that one expects that
\begin{itemize}
\item typically such equations exhibit Cantorian spectrum; 
\item $\sigma_{pp}(H_\theta)$, the singular continuous spectrum, is
  topologically typical.
\end{itemize}
%%%
%
This has been well understood only for a few models, most
prominently, for the almost Mathieu equation when
$v(x)=2\lambda\cos(x))$.
\subsection{Lyapunov exponent}
\label{sec:lyapunov-exponent}
One of the central objects of the spectral study of the quasi-periodic
equations is the Lyapunov exponent. Recall its definition.\\
Equation~\eqref{eq:1} can be rewritten as
\begin{equation}
  \label{eq:2}
  \begin{pmatrix} \psi(n+1)\\\psi(n)
  \end{pmatrix}
  = M(n\omega+\theta) \begin{pmatrix}
    \psi(n)\\\psi(n-1)
  \end{pmatrix},\quad M(x)= \begin{pmatrix} E-v(x) & -1 \\ 1&0
  \end{pmatrix}.
\end{equation}
So the large $n$ behavior of solutions to~\eqref{eq:1} can be
characterized by the limits (when they exist)~:
\begin{gather}
  \label{eq:gamma-plus}
  \gamma^+(E,\theta)=
  \lim_{n\to+\infty}\frac1n\log\|M((n-1)\omega+\theta)\cdots
  M(\theta+\omega)\,M(\theta)\|\\
  \label{eq:gamma-minus}
  \gamma^-(E,\theta)= \lim_{n\to+\infty}\frac1n\log
  \|M^{-1}(\theta-n\omega)\cdots M^{-1}(\theta-2\omega)\,
  M^{-1}(\theta-\omega) \|
\end{gather}
Furstenberg and Kesten have proved
\begin{Th}[\cite{Cy-Fr-Ki-Si:87}]
  Fix $E$. For almost every $\theta$, these limits exist, coincide and
  do not depend on $\theta$.
\end{Th}
\noindent For energies $E$ such that the limits exist, coincide and do
not depend on $\theta$, their common value is called the {\it Lyapunov
  exponent\/}; we denote it by $\gamma(E)$.
\\
We are interested in the pointwise (in both $E$ and $\theta$)
existence of the limits $\gamma^+(E,\theta)$ and $\gamma^-(E,\theta)$.
We call them the {\it right} and {\it left} Lyapunov exponents.
Speaking about the pointwise existence of the Lyapunov exponent
itself, we say that it does not exist for a pair $(E,\theta)$ when
either at least one of $\gamma^\pm(E,\theta)$ does not exist or both
of them exist, but at least one of them differs from $\gamma(E)$.
\subsection{Lyapunov exponents and the spectrum}
\label{sec:lyap-expon-spectr}
For $\omega\not\in\Q$, one has the following theorem by Ishii - Pastur
- Kotani
\begin{Th}[\cite{Cy-Fr-Ki-Si:87}]
  The absolutely continuous spectrum, $\sigma_{ac}(H_\theta),$ is the
  essential closure of the set of energies where the Lyapunov exponent
  vanishes.
\end{Th}
\noindent This theorem immediately implies
\begin{Cor}[\cite{Cy-Fr-Ki-Si:87}]
  \label{cor:1}
  If $\gamma(E)$ is positive on $I$, an interval, then the spectrum in
  $I$ (if any) is singular, $\sigma\cap I\subset\sigma_s$.
\end{Cor}
\noindent As, in general, singular continuous spectrum can be present,
in this statement, one cannot replace $\sigma_s$, the singular
spectrum, with $\sigma_{pp}$, the pure point spectrum. One may ask if
it is possible to characterize the singular continuous spectrum in
terms of the Lyapunov exponent. \\
Consider equation~(\ref{eq:1}) on the interval $E\in I$ where
$\gamma(E)>0$.\\
Almost surely, for a given $\theta$, the Lyapunov exponents exist a
priori only almost everywhere in $E$.  Denote by $I_{\text{ Lyapunov
  }}$ the subset of $I$ where
$\gamma^+(E,\theta)$
and $\gamma^-(E,\theta)$ both exist and are positive. For $E\in
I_{\text{ Lyapunov }} $, the solutions to~\eqref{eq:1} have to
increase or decrease exponentially (see,
e.g.,~\cite{Cy-Fr-Ki-Si:87}).\\
This implies that the singular continuous component of the spectral
measure vanishes on $ I_{\text{ Lyapunov }}$. So, it can be positive
only on $I\setminus I_{\text{ Lyapunov }}$.  And, the latter must
happen if the spectrum on $I$ is singular continuous.
\subsection{B. Simon's example:}
\label{sec:b.-simons-example}
We now recall an example by B.Simon showing that, for quasi-periodic
operators, one can find singular continuous spectrum on an interval
where the Lyapunov exponent is positive.\\
Consider the Almost Mathieu equation, i.e., equation~(\ref{eq:1}) with
$v(\theta)=2\lambda\cos\theta$.
\\
For this equation, by Herman's theorem (\cite{MR85g:58057}),
$\gamma(E)\ge\log \lambda$.  We assume that $\lambda>1$.  Then,
$\gamma(E)$ is positive for all $E$, and the spectrum is singular.
\\
Let the frequency $\omega$ be such that, for some infinite sequence $(p_m,q_m)
\in\N\times\N^*$,
\begin{equation*}
  \left|\omega-\frac{p_m}{q_m}    \right|\leq m^{-q_m}.
\end{equation*}
Such Liouvillean frequencies are topologically typical but of zero
measure.
\\
One has
\begin{Th}[\cite{Cy-Fr-Ki-Si:87}]
  \label{thr:2}
  Under the above conditions, there are no eigenvalues and the
  spectrum is purely singular continuous.
\end{Th}
\noindent Note that this result is a consequence of a theorem by A.
Gordon (see~\cite{MR878048,MR0458247}) which roughly says that when
the quasi-periodic potential can be super-exponentially well
approximated by periodic potentials, the equation~\eqref{eq:1} does
not admit any decreasing solutions.
\\
Note that actually, in the case of the almost Mathieu equation,
Gordon's result implies that any of its solution $\phi$ satisfies the
inequality
\begin{gather*}
  \overline{\lim}_{m\to\infty} \max\left(\,\phi(\pm
    q_m),\phi(\pm2q_m)\,\right)\geq \frac12 \phi(0),\\
  \phi(n)=(|\psi(n+1)|^2+|\psi(n)|^2)^{1/2}.
\end{gather*}
This means that the corresponding generalized eigenfunctions have to
have infinitely many humps located at some of the points $\pm q_m$,
$\pm2q_m$, \ $m\in\N$. These humps prevent the solutions from being
square summable.
\subsection{Non-trivial model problem}
\label{sec:non-trivial-model}
In the present note, we concentrate on the model equation
\begin{gather}
  \label{eq:harper}
  \psi(n+1)+\psi(n-1)=\lambda v_0(n\omega+\theta)\,\,\psi(n),\quad
  n\in\Z,\\
  \label{eq:v0}
  v_0(\theta)=2\,e^{i\pi\omega/2}\sin(\pi\theta).
\end{gather}
where $0<\omega<1$ is an irrational frequency, $1<\lambda$ is a
coupling constant, and $\theta$ is the ergodic parameter. Actually, up
to a shift in $\theta$, this is an Almost Mathieu equation with the
spectral parameter equal to zero.\\\\
We study this equation for the following reasons:
\begin{enumerate}
\item a large part of analysis is quite simple whereas (we believe
  that) to carry it out one has to use a non trivial renormalization
  procedure;
\item the techniques developed in this study can be generalized to the
  case of real analytic potentials $v$;
\item this model is related to various self-adjoint models via a
  cocycle representation (see~\cite{MR2182060}), e.g., it comes up
  naturally when studying the spectral properties of the equation
  \begin{gather*}
    -\psi''(t)+\alpha \sum_{l\ge 0}
    \delta\left(\,l(l-1)/2+l\phi_1+\phi_2\,-\,t\,\right)\,\psi(t)=E\psi(t).
  \end{gather*}
\end{enumerate}
For the model equation~(\ref{eq:harper}), our ultimate goal is to
describe the set of $\theta$ for which the Lyapunov exponent exists or
does not exist and to describe the solutions both when the Lyapunov
exponent  exists and does not exist.\\\\
We concentrate on the case of frequencies complementary to the
frequencies occurring Simon's example. And, in the case when the
Lyapunov exponent does not exist, this leads to a new scenario
for the behavior of  solutions of~(\ref{eq:1}).\\\\
Our main tool is the the {\it monodromization} renormalization method
introduced by V. Buslaev - A. Fedotov originally for the
semi-classical study of the geometry of the spectrum of one
dimensional finite difference almost periodic equations,
see~\cite{MR1300917}. The idea was to construct Weyl solutions outside
the spectrum but, at each step of the renormalization, closer to
spectrum so as to uncover smaller and smaller gaps in the spectrum.
Now, essentially, we use it to study the solutions of the model
equation on the spectrum.
\section{ Existence of the Lyapunov exponent for the model equation}
\label{sec:exist-lyap-expon}
We now formulate our results on the pointwise existence of the right
Lyapunov exponent $\gamma^+(\theta)$ for the model
equation~(\ref{eq:harper}); as we have set the energy parameter, to a
fixed value, we omit it in the Lyapunov exponents. The right Lyapunov
exponent is defined by the formula~(\ref{eq:gamma-plus}) with
\begin{equation}
  \label{eq:M-cocycle}
  M(x)= \begin{pmatrix} \lambda v_0(\theta) & -1 \\ 1&0\end{pmatrix},
\end{equation}
where $v_0$ is given by~(\ref{eq:v0}).\\
Note that for $\gamma^-(\theta)$, the left Lyapunov exponent, one has
similar results.
\subsection{Main result}
\label{sec:main-result}
Here, we formulate a sufficient condition for the existence of the
Lyapunov exponent. Therefore, we need to introduce some notations.\\
For $L=0,1,2\dots$, define
\begin{equation*}
  \omega_{L+1}=\left\{\,\frac1{\omega_L}\,\right\},\quad
  \omega_0=\omega.
\end{equation*}
where $\{a\}$ is the fractional part of $a\in\R$, and
\begin{equation*}
  \lambda_{L+1}=\lambda_L^{\frac1{\omega_L}},\quad \lambda_0=\lambda
\end{equation*}
\begin{Rem} {\it The numbers $\{\omega_l\}_{l=1}^\infty$ are related
    to the continued fraction expansion of $\omega$:
\begin{equation*}
 \omega= \frac{1}{\dsize a_1+\omega_1}=
 \frac1{\dsize a_1+\frac1{\dsize a_2+\omega_2}}=
 \frac1{\dsize a_1+\frac1{\dsize a_2+\frac1{\dsize
       a_3+\omega_3}}}=...= \frac1{\dsize a_1+\frac1{\dsize a_2+\frac
1{\dsize a_3+\frac1{\dsize a_4+\dots}}}}
\end{equation*}
where $a_1,a_2,a_3\dots\,\in\N$ are the elements of the continuous
fraction for $\omega$.\\
It is well known that, for any $l\in N$, one has
$\omega_l\omega_{l+1}\le 1/2$. This implies that the numbers
$\lambda_l$ increase super-exponentially.}
\end{Rem}
Furthermore, for a given $\omega\in (0,1)\setminus \Q$ and $s\in
(0,1)$, define the following sequence
\begin{equation}
  \label{eq:sL}
  s_L=\left\{\frac{s_{L-1}}{\omega_{L-1}}\right\},\quad s_0=s.
\end{equation}
One has
\begin{Le}
  \label{le:1}
  If $s=k_0+\omega_0 l_0$, where $k_0,\,l_0\in \Z$, then
  \begin{itemize}
  \item for all $L$, one has $s_L= k_L+\omega_L l_L$ with
    $k_L,\,l_L\in \Z$;
  \item if $k_0>0$, then the sequence $(k_{2L})_{L\geq0}$ is
    monotonically decreasing until it vanishes and then it stays
    constant equal to 0;
  \item let $k_0>0$ and $L$ be the first number for which $k_{2L}=0$,
    then
    \begin{equation*}
      k_0\omega_0\omega_1\dots\omega_{2L-1} \le 2.
    \end{equation*}
  \end{itemize}
\end{Le}
\noindent For a given $L\in\N$, define $K(2L,\omega)$ being the
maximal $k_0$ such that $k_{2L}=0$ and set
$K(2L-1,\omega)=K(2L,\omega)$.\\
Now, we are ready to discuss the Lyapunov exponent.  We have
\begin{Th}
  \label{thr:1}
  Pick $\lambda>1$ and $\omega\in(0,1)$ irrational. Assume that there
  exists a function $M:\N\to\N$ such that $M(L)< L$ and that, for
  $L\to\infty$,
  \begin{equation}\label{eq:M-L}
    \omega_{M(L)}\omega_{M(L)+1}\dots \omega_{L-1}\to 0,\quad{\rm and}\quad
    \lambda_{M(L)}\omega_{M(L)}\omega_{M(L)+1}\dots \omega_{L }\to \infty.
  \end{equation}
  For a given $0\le \theta\le 1$, the Lyapunov exponent
  $\gamma^+(\theta)$ for equation~\eqref{eq:harper} exists if, for all
  sufficiently large $L$, one has:
  \begin{equation}
    \label{eq:good-theta}
    \left| \theta-k-l\omega_0\right|\ge 
    \omega_0\omega_1\dots\omega_{M(L)-1}
    e^{-\frac{1}{\omega_0\omega_1\cdots\omega_{M(L)-1}}},
  \end{equation}
  for all $k,l\in\Z$ such that $0\le k+l\omega_0\le 1$ and
  \begin{equation}\label{eq:k-for-L}
    K(M(L),\omega)<  k\le K(\tilde
    L,\omega), \quad\quad \tilde L=\left\{\begin{array}{cc} L& \text{ if
        } L \text{ is even},\\ L+1 & \text{ otherwise}.\end{array}\right.
  \end{equation}
  Furthermore, when $\gamma^+(\theta)$ exits, it is equal to
  $\log\lambda$.
\end{Th}
\noindent One also has a similar statement on the pointwise existence
of the left Lyapunov exponent $\gamma^-$.  Note that for $\gamma^-$ to
exist, $\theta$ has to avoid neighborhoods of the points $k+l\omega_0$
with negative $k$. \\
Now, turn to a discussion of the results given in Theorem~\ref{thr:1}.
\subsection{Admissible frequencies}
\label{sec:admiss-freq}
Denote by $\Omega$ the set of $\omega\in(0,1)$ satisfying the
conditions of Theorem~\ref{thr:1}
\subsubsection{The measure of $\Omega$}
\label{sec:measure-omega}
Khinchin's famous result (see e.g.~\cite{MR98c:11008}) on the
geometric means of the products of the elements of the continued
fractions implies
\begin{Le}
  \label{le:2} 
  ${\rm mes}\,\, \Omega=1$.
\end{Le}
\noindent\demo Let $\{a_l\}$ be the elements of the continued
fraction for $\omega$. By Khinchin, for almost all $\omega$, one has
$\lim_{L\to\infty} \left(a_1a_2\dots a_L\right)^{\frac1L}=C$, where
$C=2,6...$ is a universal constant.  Pick $l\in \N$. One has
$\frac1{2a_l}<\omega_{l-1}<\frac1{a_l}$. Therefore, for almost all
frequencies $\omega$,
\begin{equation*}
  \overline{{\rm lim}}_{L\to\infty} 
  \left(\omega_0\omega_1\dots\omega_{L-1}\right)^{\frac1L}\le
  \frac1C,\quad\quad
  \underline{{\rm lim}}_{L\to\infty} 
  \left(\omega_0\omega_1\dots\omega_{L-1}\right)^{\frac1L}\ge
  \frac1{2C}.
\end{equation*}
Such $\omega$ belong to $\Omega$: in~(\ref{eq:M-L}) one can take
$M(L)=[L/2\,]$.\qed
\subsubsection{Liouvillean numbers in $\Omega$}
\label{sec:liouv-numb-omega}
Recall that an irrational number $\omega$ is called Liouvillean if,
for any $n\in\N$, there are infinitely many $(p,q)\in\Z\times\N$ such
that
\begin{equation*}
  \left|\omega-\frac{p}{q}\right|\le \frac1{q^n}. 
\end{equation*}
(see e.g.\cite{MR98c:11008}).\\
One has
\begin{Le}
  \label{le:3}
  The set $\Omega$ contains Liouvillean numbers satisfying
  \begin{equation}\label{eq:omega-Liouville-example}
    \left|\omega-\frac{p}{q}\right|\le \frac1{q\lambda^{cq}},\quad\quad 
    c=c(\omega)>0,
  \end{equation}
  for infinitely many $(p,q)\in\Z\times\N$.
\end{Le}
\noindent\demo We construct a Liouvillean $\omega\in\Omega$ by
choosing inductively $(a_l)_{l\geq1}$, the elements of its continued
fraction.  Therefore, we pick $a_1\geq1$ large and, for all $L\ge 1$,
we choose $a_{L+1}$ so that
\begin{equation}\label{eq:aL}
  \frac12\, a_{L+1}\le (a_1a_2\dots a_L)^{-1}\lambda^{a_1a_2\dots a_{L}}\le
  a_{L+1}.
\end{equation}
We now check that such an $\omega$ belongs to $\Omega$. Therefore, we
check that one has~(\ref{eq:M-L}) for $M(l)=l-1$.  As $\lambda>1$, the
sequence $(a_{l})_l$ is quickly increasing, and so
\begin{equation}\label{eq:M-l-1}
  \omega_{l-1}\to 0,\quad l\to\infty.
\end{equation}
Furthermore, as, for all $l\ge 0$, one has
$\omega_{l}=(a_{l+1}+\omega_{l+1})^{-1}$, we get
\begin{equation*}
  \omega_{l-1}\omega_l\lambda_{l-1}>\frac1{4a_la_{l+1}}\,\lambda^{a_1a_2\dots
    a_{l}}\ge \frac{a_1a_2\dots a_{l-1}}{8}.
\end{equation*}
This implies that
\begin{equation*}
  \lambda_{l-1}\omega_{l-1}\omega_l\to\infty,
\end{equation*}
and so $\omega\in\Omega$.\\
Now, let us check that $\omega$
satisfies~(\ref{eq:omega-Liouville-example}) (and, thus, is a
Liouville number).  Consider $\left(\frac{p_l}{q_l}\right)$, the
sequence of the best approximates for $\omega$.  Recall that (see
e.g.\cite{MR98c:11008}), for all $l\in \N$,
\begin{gather}\label{eq:qp-1}
  \left|\omega-\frac{p_l}{q_l}\right|\le \frac1{a_{l+1}q_{l}^2},\\
  \nonumber\\
  \label{eq:qp-2}
  q_{l+1}=a_{l+1}q_l+q_{l-1},\quad\quad q_1=a_1,\quad q_0=1.
\end{gather}
The relations~\eqref{eq:qp-2} imply that
\begin{equation}\label{eq:a-q}
  a_{l}a_{l}\dots a_2a_1< q_{l}< P\,a_{l}\dots a_2a_1,\quad
  P=\prod_{l=1}^\infty\left(1+\frac1{a_la_{l+1}}\right);
\end{equation}
the product $P$ converges as the sequence $(a_l)_l$ is quickly
increasing.  Relations (\ref{eq:a-q}) and~(\ref{eq:aL}) imply that
$a_{l+1}\ge q_l^{-1}\,\lambda^{q_l/P}$. This and~(\ref{eq:qp-1})
imply~(\ref{eq:omega-Liouville-example}).\qed
\subsection{The set of ``bad'' phases}
\label{sec:set-bad-theta}
For given $\lambda>1$ and $\omega\in \Omega$, denote by $\Theta$ the
set of phases $\theta$ not satisfying (\ref{eq:good-theta}) for
infinetly many $L$. One has
\begin{Le}
  The set $\Theta$ is topologically typical (countable intersection of
  dense open sets) and, under the condition
  \begin{equation}\label{eq:good-theta0}
    \sum_{L>0}^\infty
    \left(\lambda_{M(L)}\omega_{M(L)}\omega_{M(L)+1}\dots \omega_{L
      }\right)^{-1}<\infty
  \end{equation}
  (which is stronger than~(\ref{eq:M-L})), it has zero Lebesgue
  measure.
\end{Le}
\noindent\demo For a given $L>0$, denote the set of $\theta$ not
verifying~(\ref{eq:good-theta}) by $\Theta_L$.  Then
\begin{equation}
  \label{eq:cupcap}
  \Theta=\bigcap_{N\ge 0}\,\,\bigcup_{L\ge N} \Theta_L.
\end{equation}
Thus, $\Theta$ is a countable intersection of open sets.\\
As $\omega$ is irrational, the points $\theta_{k,l}=k+\omega_0l$
($k,l\in\Z$, $k\ge 0$) are dense in the interval $(0,1)$. So, to
complete the proof of the first property of $\Theta$, it suffices to
show that the set $\bigcup_{L\le N} \Theta_L$ contains all the points
$\theta_{k,l}$ with $k$ sufficiently large. But, this
follows from~(\ref{eq:k-for-L}) and the inequality $M(L)<l$.\\
Finally, note that, by~(\ref{eq:good-theta}),
\begin{equation*}
  {\rm mes}\,\Theta_L\le \frac1\omega\,
  K(L)\omega_0\omega_1\dots\omega_{M(L)-1} 
  \lambda_M(L)^{-1}\le \frac2\omega\,\left(\lambda_M(L)
    \omega_{M(L)}\dots\omega_{L-1}\omega_L\right)^{-1}.
\end{equation*}
Under the condition~(\ref{eq:good-theta0}), this implies that the
Lebesgue measure of $\Theta$ is zero.\qed
\subsection{Heuristics and the statement of Theorem~\ref{thr:1}}
\label{sec:heur-stat-theor}
Let us now describe some heuristics ``explaining'' Theorem~\ref{thr:1}.\\\\
Consider a continuous version of equation~(\ref{eq:harper}) :
\begin{equation}
  \label{eq:harper-continuous}
  \phi(s+\omega)+\phi(s-\omega)=\lambda v_0(s)\,\,\phi(s),\quad s\in\R.
\end{equation}
If $\phi$ satisfies this equation, then the formula
$\psi(n)=\phi(n\omega+\theta)$, \ $n\in\Z$, defines a solution
to~(\ref{eq:harper}).\\
If $\lambda>>1$, then one can expect that, on a fixed compact interval,
equation~(\ref{eq:harper-continuous}) has an exponentially increasing
solution $\phi^+$ with the leading term $\phi^+_0$ satisfying the
equation
\begin{equation}
  \label{eq:half-of-harper-continuous}
  \phi^+_0(s+\omega)=\lambda v_0(s)\,\,\phi^+_0(s),\quad s\in\R.
\end{equation}
For the last equation, one can easily construct a solution $\phi_0^+$
that is analytic and has no zeros in the band $0<{\rm re}\,
s<1+\omega$.  One can extend this solution analytically to the left of
this band using equation~(\ref{eq:half-of-harper-continuous}). As
$v_0$ vanishes at integers, $\phi^+_0$ has zeros at all the points of
the form $s_{k,l}= k+l\omega$ where $k,\,l>0$ are integers.\\
If there is a true solution to~(\ref{eq:harper-continuous}) with the
leading term $\phi_0^+$, then~(\ref{eq:harper}) has a solution
$\psi^+$ with the leading term $\phi_0^+(n\omega+\theta)$.
Furthermore, if $\theta\in (0,1)$ admits the representation
$\theta=k_0-l_0\omega $ with some positive integers $k_0$ and $l_0$,
then, at least for sufficiently large $\lambda$, the leading term of
$\psi^+$ increases exponentially on the ``interval'' where
$-\frac{k_0}{\omega}+l_0<n<l_0+1$ and then vanishes at the points
$n=l_0+1,\,l_0+2,\dots$.\\
The equality $\theta=k_0-l_0\omega $ can be interpreted as a
quantization condition: when this condition is satisfied, the solution
$\psi^+$ that is exponentially growing up to the point $n=l_0$, at
this point, changes to the exponential decay.\\
So, it is natural to expect that the solution $\psi^+$ keeps growing
up to the infinity if $\theta$ is ``far enough'' from all the points
of the form $k_0-l_0\omega$ with positive integers $k_0$ and $l_0$.
Hence, the right Lyapunov exponent should exist.
\section{Non-existence of the Lyapunov exponent}
\label{sec:non-exist-lyap}
Theorem~\ref{thr:1} is rather rough in the sense that the sizes of the
``secure intervals'' that $\theta$ has to avoid for the Lyapunov
exponent to exist (see~(\ref{eq:good-theta})) are too big. This is
actually due to the fact that, under the conditions of
Theorem~\ref{thr:1}, one has much more than the existence of Lyapunov
exponent. Roughly, under these conditions, for each $L$ large enough,
equation~(\ref{eq:harper}) has solutions that, locally, on intervals
of length of order $(\omega_0\omega_1\dots \omega_{M(L)-1})^{-1}$, can
have complicated behavior whereas globally, on the interval
$0<k<{\mathcal K}$ of length of order $(\omega_0\omega_1\dots
\omega_{L})^{-1}$, they are nicely exponentially increasing.\\\\
Our method also allows a precise description of the set of $\theta$
where the Lyapunov exponent does not exist. The structure of this set
is quite complicated; in the present note, we only describe it for
frequencies in $\Omega_1\subset\Omega$, the set of $\omega$ satisfying
the conditions
\begin{equation}\label{eq:M-L-2}
  \omega_{L}\to 0,\quad{\rm and}\quad
  \lambda_{L-1}\omega_{L-1}\omega_{L }\to \infty.
\end{equation}
instead of~(\ref{eq:M-L}).\\
One has the following two statements:
\begin{Th}
  \label{le:non-exist-even} 
  Pick $\lambda>1$. Let $\omega\in\Omega_1$. For a $0<\theta<1$,
  define the sequence $\{s_L\}$ by~(\ref{eq:sL}) with $s_0=\theta$.
  Assume that there is a positive constant $c$ such that for
  infinitely many even positive integers $L$ one has
  \begin{equation*}
    {\rm dist}\,\left(\,s_{L-1}\,,\,\omega_{L-1}\cdot \N\,\right)\le
    \omega_{L-1}\,\lambda_L^{-c}\quad{\rm and}\quad s_{L-1}\ge c.
  \end{equation*}
  Then, the right Lyapunov exponent $\gamma^+(\theta)$ does not exist.
\end{Th}
\noindent and
\begin{Th}
  \label{le:non-exist-odd}
  Pick $\lambda>1$. Let $\omega\in\Omega_1$. For a $0<\theta<1$,
  define the sequence $\{s_L\}$ by~(\ref{eq:sL}) with $s_0=\theta$.
  Assume that there is a positive constants $c$ and $N$ such that for
  infinitely many odd positive integers $L$ one has
  \begin{equation*}
    {\rm dist}\,\left(\,s_{L}\,,\,\omega_{L}\cdot \N\,\right)\le
    \lambda_L^{-c},\quad s_{L-1}\le 1-c,
    \quad{\rm and}\quad s_{L}\le \omega_L\,N.
  \end{equation*}
  Then, the right Lyapunov exponent $\gamma^+(\theta)$ does not exist.
\end{Th}
\noindent The above two theorems are sharp: in the case of
$\omega\in\Omega_1$, if the Lyapunov exponent does not exist, then
$\theta$  satisfies the conditions of one of the them. \\\\
As for the behavior of the solutions, in both cases, roughly, we find
that, for infinitely many $L$, even if we forget of the complicated
local behavior of the solutions on the intervals of the length of
order $(\omega_0\omega_1\dots \omega_{M(L)-1})^{-1}$, one can see that
globally, on the interval $0<k<{\mathcal K}$ of the length of order
$(\omega_0\omega_1\dots \omega_{L})^{-1}$, the solutions change from
exponential growth to the exponential decay. For example, in the case
of Theorem~\ref{le:non-exist-odd}, there exists solutions that, at
first, are globally exponentially increasing then are globally
exponentially decaying, the length of the interval of increase and the
interval of decrease being of the same order. Here we use the word
globally to refer to the fact that this exponential growth or decay
happens at a large scale.
\section{The main ideas of the proof}
\label{sec:main-ideas-proof}
As we have mentioned in the introduction, the main tool of the proof
is the {\it monodromization} renormalization method. The new idea is
that one can consider the infinite sequence of the almost periodic
equations arising in the course of the monodromization as a sequence
of equations describing a given solution of the input equation on
larger and larger intervals, the ratio of their length being
determined by the continued fraction of the frequency.\\
Now, the renormalization formulas can be written in the form
\begin{equation}\label{eq:renrmalization}
  \begin{split}
    M(&\theta+(k-1)\omega)\dots
    M(\theta+\omega)\,M(\theta)\\
    &\sim \Psi( \{k\omega+\theta\})\,\left[\,
      M_1(\theta_1-\omega_1)\,M_1(\theta-2\omega_1)\dots
      M_1(\theta-k_1\omega_1)\,\right]^t\,\Psi^{-1}(\theta).
  \end{split}
\end{equation}
Here, ``$\sim$'' means ``equal up to a sign'',
\begin{itemize}
\item $M(\theta)=\begin{pmatrix} 2\lambda \sin(\pi\theta) &
    -e^{-i\pi\theta}\\ e^{i\pi\theta}& 0\end{pmatrix}$; the second
  order difference equation~(\ref{eq:harper-continuous}), the
  continuous analog of~(\ref{eq:harper}), is equivalent to the first
  order matrix difference equation
  \begin{equation}\label{eq:matrix-eq}
    \Psi(s+\omega)=M(s)\Psi(s),\quad s\in \R;
  \end{equation}
\item $\Psi$ is a fundamental solution to~(\ref{eq:matrix-eq}), i.e.,
  such that $\Psi(s)\in SL(2,\C)$ for all $s$;
\item ${}^t$ denotes the transposition;
\item $M_1$ is a monodromy matrix corresponding to this solution,
  i.e., the matrix defined by $\Psi(s+1)=\Psi(s) M_1^t(s/\omega)$.
\end{itemize}
The new constants $\omega_1$, $\theta_1$ and the number $k_1$ are
defined by
\begin{equation*}
  \omega_1=\{\,1/\omega\,\},\quad \theta_1=\{\,\theta/\omega\,\},\quad
  k_1=[\,\theta+k\omega\,].
\end{equation*}
And, as usual $\{a\}$ and $[a]$ denote the fractional and the integer
part of $a\in\R$. \\
Formula~\eqref{eq:renrmalization} relates the study of the matrix
product $M(\theta+(k-1)\omega)\dots M(\theta+\omega)\,M(\theta)$ to
that of a similar product: the monodromy matrix $M_1$ is unimodular
and, as the matrix $M$, it is 1 anti-periodic.  One can apply the same
renormalization formula for the new matrix product and so on.  It is
easy to check that after a finite number of renormalizations, one gets
a matrix product containing at most ... one matrix.  This feature
recalls the renormalization of the quadratic exponential sums
carried out by Hardy and Littlewood (see~\cite{Ha-Li:14,MR2182060}).\\
At each step of the monodromization, one has to study similar
difference equations $\Psi_L(s+\omega_k)=M_L(s)\Psi_L(s)$, \
$L=0,1,2,\dots$.  One needs to have a good enough control of their
solutions but only on
one fixed compact interval namely $[0,1]$.\\
For our model, one can choose the fundamental solutions so that all
the matrices $M_L$ have the same functional structure, and the numbers
$(\lambda_L)$ are the successive coupling constants in these
equations.\\
For $\lambda=\lambda_0>1$, the sequence $(\lambda_L)_L$ tends to
infinity very rapidly; this enables an effective asymptotic analysis
of the successive equations. For general almost periodic equations, one
finds an analogous effect at least when the coupling constant in the
input equation is large enough.
\def\cprime{$'$} \def\cydot{\leavevmode\raise.4ex\hbox{.}}

%
% \bibliographystyle{plain}
% \bibliography{biblio}
%
% \end{thebibliography}
%
\end{document}